\documentclass[reprint,prb,twocolumn,showpacs,superscriptaddress,aps,longbibliography,floatfix]{revtex4-2}
\usepackage[utf8]{inputenc}
\usepackage{graphicx}
\usepackage[colorlinks=true,citecolor=blue]{hyperref}
\usepackage{amsmath}
\usepackage{amssymb}
\usepackage{soul}
\usepackage{gensymb}
\usepackage{amsfonts,amssymb,amsbsy}
\usepackage{siunitx}
\usepackage{float}
\usepackage{physics}
\usepackage{booktabs}
\usepackage[percent]{overpic}

\usepackage{xcolor}
\usepackage{natbib}

\usepackage{lineno}
\begin{document}

\title{Multiferroic Quantum Dot in an Artificial van der Waals  Heterostructure}

\author{Antti Karjasilta}
\affiliation{Aalto University, Department of Applied Physics, 00076 Aalto, Finland}

\author{Mohammad Amini}
\affiliation{Aalto University, Department of Applied Physics, 00076 Aalto, Finland}

\author{Liwei Jing}
\affiliation{University of Jyväskylä, Department of Physics, FI-40014 University of Jyväskylä, Finland}

\author{Robert Drost}
\affiliation{Aalto University, Department of Applied Physics, 00076 Aalto, Finland}

\author{Jose Lado}
\affiliation{Aalto University, Department of Applied Physics, 00076 Aalto, Finland}

\author{Shawulienu Kezilebieke}
\affiliation{University of Jyväskylä, Department of Chemistry, 
FI-40014 University of Jyväskylä, Finland}

\author{Peter Liljeroth}
\email{Corresponding authors. Emails: peter.liljeroth@aalto.fi, adolfo.oterofumega@aalto.fi}
\affiliation{Aalto University, Department of Applied Physics, 00076 Aalto, Finland}

\author{Adolfo O. Fumega}
\email{Corresponding authors. Emails: peter.liljeroth@aalto.fi, adolfo.oterofumega@aalto.fi}
\affiliation{Aalto University, Department of Applied Physics, 00076 Aalto, Finland}

\begin{abstract}
Quantum dots (QDs) provide a versatile platform for engineering quantum-confined electronic states with functionalities relevant for optoelectronics, spintronics, and quantum technologies. While substantial progress has been achieved in coupling confined states to spin, valley,  topological, or ferroelectric degrees of freedom, the realization of a multiferroic QD in which quantum confinement simultaneously intertwines with magnetism and ferroelectricity remains elusive. Here, we engineer a multiferroic QD in an artificial van der Waals heterostructure grown by molecular beam epitaxy under ultra-high-vacuum conditions. The heterostructure consists of ferroelectric SnTe nanoislands deposited on the layered magnet CrBr$_2$ supported on highly oriented pyrolytic graphite. Combining scanning tunneling microscopy and spectroscopy with \textit{ab initio} calculations and low-energy tight-binding models, we demonstrate the emergence of spin-polarized discretized electronic states confined within the SnTe islands. Remarkably, the spectroscopic response of the QD strongly depends on the ferroelectric domain configuration of the SnTe nanoislands, demonstrating an interplay between quantum confinement, magnetic exchange, and ferroelectric order at the atomic scale. Our results establish engineered van der Waals heterostructures as a platform for multiferroic quantum confinement and open new routes toward electrically tunable quantum spintronic devices.

\end{abstract}

\date{\today}

\maketitle

\section{Introduction}

Quantum dots (QDs) are nanoscale systems in which charge carriers are spatially confined, giving rise to discrete, atomic-like electronic states whose energies can be engineered through size, shape, composition, electrostatic gating, strain, and the surrounding dielectric environment.\cite{Alivisatos1996,Klimov2000,Flsch2014} This tunability has made QDs central building blocks for optoelectronics, photovoltaics, quantum light sources, spin qubits, sensing, and nanoscale electronics.\cite{Rauch2009,Cho2009,GarcadeArquer2021,Kim2022,RevModPhys.95.025003} 
Spectroscopy techniques have enabled the characterization of the QDs' energy levels.\cite{PhysRevLett.68.3088,PhysRevLett.73.2252} 
Specifically, scanning tunneling microscopy (STM) and atomic force microscopy (AFM) have allowed the determination of the real-space structure of the confined modes.\cite{Crommie1993,Banin1999,Jdira2008,Woodside2002,PhysRevLett.89.086801,PhysRevLett.94.056802,Cockins2010,PhysRevLett.115.026101,Mohn2012,Peng2021}
Beyond conventional semiconductor QDs, increasing effort has focused on QDs in which additional internal degrees of freedom, such as spin,\cite{Awschalom2013} valley,\cite{PhysRevLett.108.126804,Garreis2024} topology,\cite{Kiczynski2022}  or ferroelectric polarization,\cite{Long2025} can be coupled to the confined electronic spectrum, providing promising functional platforms for  spintronics and quantum computing.\cite{PhysRevA.57.120,RevModPhys.79.1217} 
However, realizing a quantum dot in which confined electronic levels are simultaneously coupled to magnetic order and ferroelectric domains remains an open route toward electrically tunable, quantum-confined spintronic states.

Van der Waals (vdW) materials and heterostructures provide an exceptional platform for engineering emergent quantum phenomena through the controlled combination of distinct atomically thin crystals, such as magnets,\cite{Lee2016,Gong2017,Huang2017} ferroelectrics,\cite{Chang2016,Cui2018} or multiferroics.\cite{Song2022,Amini2024} 
The weak interlayer bonding and atomically sharp interfaces enable the realization of designer quantum matter.\cite{Geim2013,Novoselov2016}
This approach has led to the discovery of a wide range of emergent phenomena, including correlated insulating states and unconventional superconductivity in twisted bilayer graphene,\cite{Cao2018_ci,Cao2018_sc} artificial heavy-fermion behavior,\cite{Vao2021} topological superconductivity,\cite{Kezilebieke2020} topological crystalline phases,\cite{Jing2026} or fractional Chern insulators.\cite{Spanton2018,Cai2023} 
Beyond collective quantum phases, vdW heterostructures have also emerged as a versatile platform for quantum confinement and artificial QD engineering, including the realization of coupled QD states and orbital hybridization between confined electronic levels.\cite{Ge2023,Zhou2024,Mao2025} 
In parallel, molecular beam epitaxy (MBE) under ultra-high-vacuum (UHV) conditions has become a powerful route for the growth of high-quality complex vdW heterostructures,\cite{Zhang2022} enabling atomically controlled synthesis in a clean environment compatible with advanced surface-sensitive probes such as scanning tunneling microscopy and spectroscopy.

\begin{figure*}[t!]
    \centering
    \includegraphics[width=\linewidth]{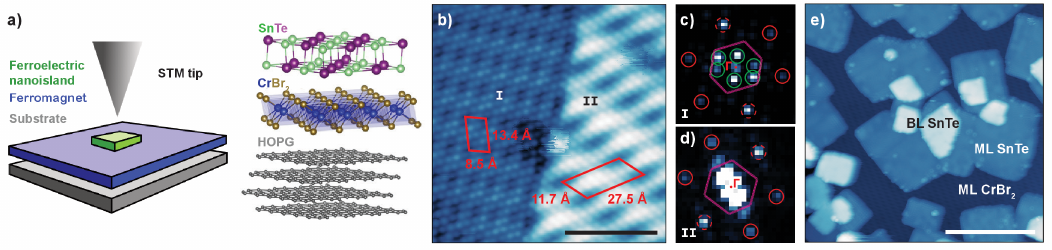}
    \caption{Growth of SnTe-CrBr$_2$ heterostructure and characterization of monolayer CrBr$_2$. \textbf{a.} Schematic illustrations of the bilayer heterostructure. \textbf{b.} Monolayer CrBr$_2$ at atomic scale inside the band gap (\SI{1.0}{\volt}, \SI{28.7}{\pico\ampere}, scale    bar \SI{4}{\nano\meter}). The unit cells of two different moiré supercells are illustrated. \textbf{c, d.} FFTs of the left (I) and right (II) domains in panel b, respectively. Red circles highlight the unit cell, with the dashed circles showing the distorted direction. The first Brillouin zone is depicted with magenta dashed lines, and in panel c the moiré superstructure is illustrated by the green circles. \textbf{e.} Full monolayer CrBr$_2$, with rectangular SnTe islands grown on top (\SI{1.2}{\volt}, \SI{8.0}{\pico\ampere}, scale bar \SI{30}{\nano\meter}).}
    \label{fig:fig1}
\end{figure*}

Here, we engineer a multiferroic QD in an artificial vdW heterostructure (Fig.~\ref{fig:fig1}a,b). Using MBE under UHV conditions, we grow a heterostructure consisting of ferroelectric SnTe nanoislands on top of the layered magnet CrBr$_2$ supported on highly oriented pyrolytic graphite (HOPG). Combining STM with \textit{ab initio} calculations and low-energy effective tight-binding models, we characterize the structural and electronic properties of this artificial multiferroic heterostructure at the atomic scale. We uncover the emergence of a multiferroic QD state in the SnTe islands, where discretized spin-polarized electronic levels coexist and intertwine with the ferroelectric order.
Furthermore, we demonstrate that the spectroscopic response of the QD is strongly modified by the ferroelectric domain configuration of the SnTe islands, revealing an interplay between quantum confinement, magnetism, and ferroelectricity in an engineered vdW platform.

\section{Results}

\subsection{Growth and characterization of SnTe / CrBr$_2$ vdW heterostructures}

Our SnTe-CrBr$_2$ heterostructure is grown on HOPG (Fig. \ref{fig:fig1}a) by MBE. We start by depositing CrBr$_2$ using a single-source deposition (details in Methods). We obtain full monolayer coverage and use atomically resolved STM images to characterize monolayer CrBr$_2$. Figure \ref{fig:fig1}b shows two distinct crystallographic domains with strongly different moiré patterns (unit cells indicated in the figure). Based on the real-space images and their corresponding Fast Fourier Transforms (FFTs) (Figs. \ref{fig:fig1}c,d), the CrBr$_2$ has a distorted triangular lattice, with the two unit vectors having lengths of 3.75$\pm$\SI{0.05}{\angstrom} and 3.99-\SI{4.02}{\angstrom} (these values are similar to the values reported in the literature for the CrBr$_2$ growth on NbSe$_2$) \cite{9lb4-hrnv}. Due to the distorted triangular lattice of the CrBr$_2$, the moiré unit cell also does not have perfect trigonal symmetry. Additionally, the moiré periodicities vary strongly depending on the angle between the HOPG substrate and the ML CrBr$_2$ as can be seen in Fig.~\ref{fig:fig1}c,d.

\begin{figure}
    \centering
    \includegraphics[width=\linewidth]{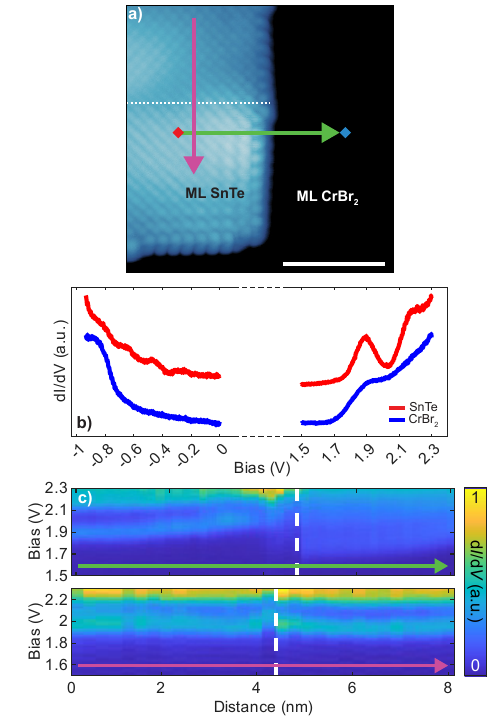}
    \caption{STS characterization of SnTe-CrBr$_2$ heterostructure. \textbf{a.} Monolayer SnTe island on monolayer CrBr$_2$. The ferroelectric distortion on SnTe is visible, with a domain boundary shown by the dashed white line. The red and blue squares show the locations of the spectra plotted in panel (b), and the green and purple arrows show the locations of the line spectra plotted in panel (c) (\SI{1.2}{\volt}, \SI{8}{\pico\ampere}, scale bar \SI{5}{\nano\meter}). \textbf{b.} d$I$/d$V$ spectroscopy of monolayer CrBr$_2$ and monolayer SnTe on top of CrBr$_2$. \textbf{c.} Line spectra from monolayer SnTe to monolayer CrBr$_2$ (top panel) and across the ferroelectric domain boundary of SnTe (bottom panel). The dashed white lines show the location of the SnTe-CrBr$_2$ edge and ferroelectric domain boundary, respectively.}
    \label{fig:fig2}
\end{figure}

After growth of CrBr$_2$, we use single-source deposition to grow SnTe. This results in rectangular SnTe monolayer islands, with minor bilayer growth (Fig. \ref{fig:fig1}e).  
We continue by confirming that SnTe monolayer remains ferroelectric when grown on top of monolayer CrBr$_2$ as this could be affected by factors such as strain or doping \cite{PhysRevLett.121.027601,Jing2026}. Previous experiments show that observing distinct ferroelectric domains and band bending near SnTe island edges are strong indicators of ferroelectricity \cite{Chang2016,Amini2023}. The ferroelectric domains can be observed directly in the STM images as shown in Fig. \ref{fig:fig2}a. This follows from the structural ferroelectricity of ML SnTe; above the ferroelectric transition temperature $T_C$, SnTe has a cubic structure, but when the temperature is lowered below $T_C$, SnTe shifts into a rhombohedral structure. The Sn and Te sublattices are displaced, which gives rise to the distinct appearance of ferroelectric domains on SnTe. 

For further characterization, we utilize scanning tunneling spectroscopy (STS) to probe the local density of states (LDOS) of the SnTe / CrBr$_2$ heterostructures. Spectroscopy on the CrBr$_2$ layer shows a band gap of  \SI{\sim2.4}{e\volt} (Fig. \ref{fig:fig2}b). The conduction band of ML SnTe grown on top of CrBr$_2$ is shifted up slightly from the values obtained on SnTe directly grown on the HOPG substrate \cite{Chang2016,Amini2023} (Figure \ref{fig:fig2}b). On HOPG, the SnTe valence band is very close to the Fermi level, indicating partial charge transfer. The valence band onset of SnTe is difficult to resolve precisely on the SnTe/CrBr$_2$ heterostructures, but our subsequent experiments (see below) suggest that the SnTe islands are charged.
Acquiring line spectra from ML SnTe to the ML CrBr$_2$ offers additional evidence of ferroelectricity in SnTe. A shift of the CB due to ferroelectric polarization can be observed on the SnTe island edges and across ferroelectric domain boundaries (Fig. \ref{fig:fig2}c) \cite{Chang2016,Amini2023}. The direction of the shift follows from the polarization direction of the SnTe. Specifically, the spectra towards the island edge shown in Fig.~\ref{fig:fig2}c shift upwards, revealing that the edge is negatively charged.

\begin{figure}
    \centering
    \includegraphics[width=\linewidth]{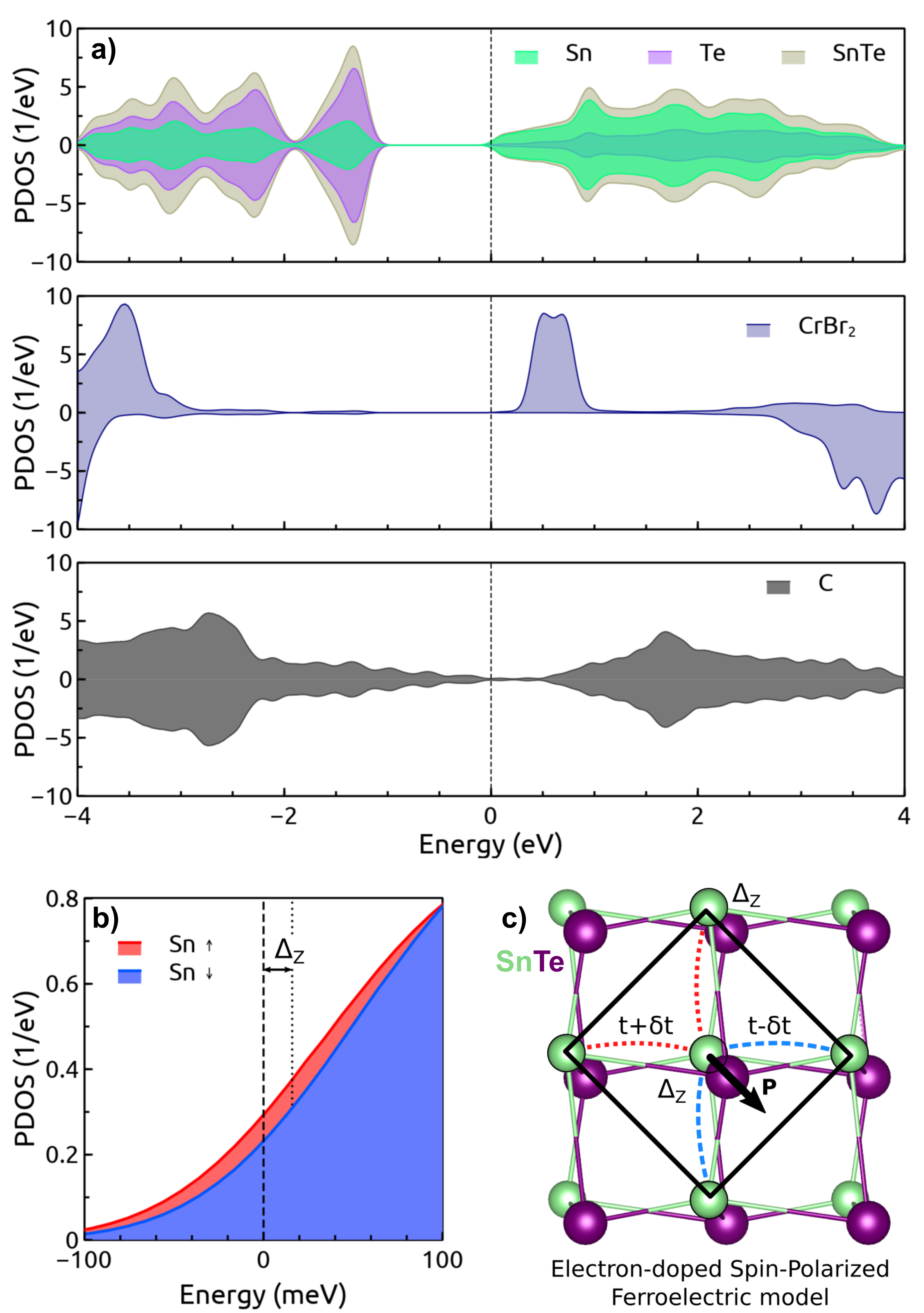}
    \caption{\emph{Ab initio} calculation on the artificial van der Waals heterostructure \textbf{a.} Projected Density of States (PDOS) for SnTe, CrBr$_2$, and graphene, respectively. Positive (negative) PDOS denotes the majority (minority) spin channel. A charge transfer between graphene and the conduction band of SnTe can be identified. \textbf{b.} SnTe PDOS comparison between the spin channels around the Fermi level. For clarity, the majority and minority PDOS are plotted as positive values. A spin-imbalance Zeeman term $\Delta_z$ is observed as a consequence of the underlying magnetic CrBr$_2$. \textbf{c.} Sketch of the low-energy minimal model that captures the electron-doped spin-polarized ferroelectric nature of SnTe in the artificial van der Waals heterostructure. }
    \label{fig:fig_theory}
\end{figure}

\begin{figure*}
    \centering
    \includegraphics[width=\linewidth]{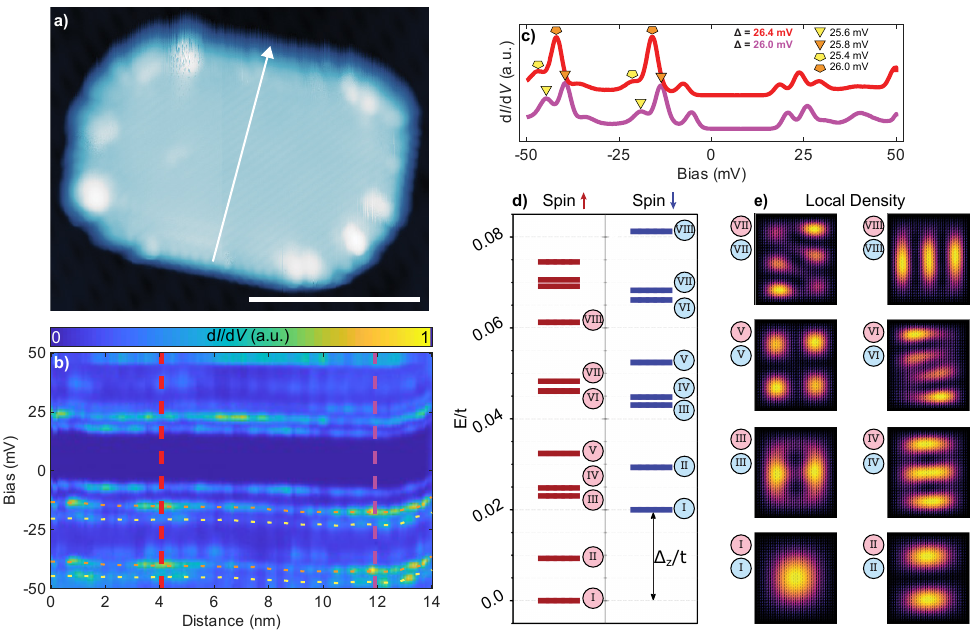}
    \caption{Low-energy LDOS features in a single ferroelectric domain SnTe island. \textbf{a.} STM image of a single ferroelectric domain SnTe island with the arrow indicating the positions of the recorded d$I$/d$V$ spectra shown in panel (b) (\SI{1}{\volt}, \SI{3}{\pico\ampere}, scale bar \SI{10}{\nano\meter}). \textbf{b.} Color-scale plot of d$I$/d$V$ spectra. The dashed orange and gray lines indicate LDOS features that have identical spatial dependence. \textbf{c.} Individual d$I$/d$V$ spectra taken from the data in panel b, indicating that the separation of doubled signals (red and orange markers) is constantly $\sim$\SI{26}{\milli\volt}. The gap size is $\sim$\SI{26}{\milli\volt}. \textbf{d, e.} Calculated energy levels (panel d) and the corresponding LDOS maps (panel e) for a spin-polarized quantum dot formed in the single-domain SnTe island.}
    \label{fig:fig4}
\end{figure*}

To gain further insight into the electronic structure of these vdW heterostructures, we performed first-principles calculations for the SnTe/CrBr$_2$/graphene stack. 
The projected density of states (PDOS) shown in Fig.~\ref{fig:fig_theory}a reveals a substantial charge transfer from graphene to the SnTe layer, resulting in an electron-doped ferroelectric SnTe subsystem. Additionally, the electronic structure of the CrBr$_2$ can be understood from an ionic point of view considering Cr$^{2+}$ embedded in the distorted octahedral environment created by Br$^{-}$ ions. This leads to a $d^4$ orbital occupation for Cr$^{2+}$. Onsite Coulomb interactions promote a high-spin state $S=2$, where the $t_{2g}$ and $d_{z^2}$ local orbitals are fully spin-polarized, leading to a magnetic insulating ground state \cite{PhysRevB.111.054403,9lb4-hrnv}. 
In the artificial vdW heterostructure, the magnetic CrBr$_2$ layer induces a spin polarization in the electronic states of SnTe through proximity exchange coupling. 
This effect becomes evident in the low-energy PDOS of SnTe, where a clear spin imbalance is observed around the Fermi level (Fig.~\ref{fig:fig_theory}b). 
The resulting spin splitting can be effectively described by an exchange-induced Zeeman term $\Delta_z$, which lifts the spin degeneracy of the electronic states in the QD while preserving the ferroelectric character of the SnTe layer. These results establish that the artificial heterostructure simultaneously hosts electron doping, ferroelectric order, and spin polarization. 
Motivated by the \emph{ab initio} calculations, we introduce a minimal low-energy model (Fig.~\ref{fig:fig_theory}c) consisting of an electron-doped ferroelectric SnTe layer subject to an effective exchange field $\Delta_z$.
For this minimal model, we consider a single orbital on the Sn sites, since the doped conduction band has a clear Sn character (Fig.~\ref{fig:fig_theory}a).
The spin polarization induced by the underlying CrBr$_2$ layer is captured through the exchange-induced Zeeman term $\Delta_z$, while the ferroelectric distortion is described by a symmetry-breaking modulation of the nearest-neighbor hopping amplitudes $t\pm\delta t$. The sign and orientation of $\delta t$ encode the ferroelectric polarization direction, allowing different ferroelectric domain configurations to be naturally incorporated within the model. This minimal description captures the key ingredients governing the low-energy electronic properties of the heterostructure and serves as the basis for the analysis of the multiferroic QD presented below.

\subsection{Single-domain multiferroic quantum dot}

Carrying out spectroscopy with higher energy resolution, we observe a multitude of features close to zero bias (Fig. \ref{fig:fig4}a,b). The spectra have a hard gap in the range from \SI{\sim10}{\milli\volt} to \SI{\sim50}{\milli\volt} depending on the exact island configuration. The most likely explanation is that this corresponds to Coulomb blockade physics \cite{Banin1999,PhysRevB.65.165334,Kouwenhoven_2001,Swart2016} and that the SnTe islands act as QDs. Following the particle-in-a-box picture, additional charge carriers in the SnTe island give rise to the discrete energy levels. When the QD is sufficiently small, the confinement effects dominate, and the separation between the discrete energy levels is determined by the size of the QD. In our experiments, the doping (the number of additional charge carriers on the SnTe QD energy levels) is expected to depend on the size of the SnTe island. This makes a direct comparison between the different SnTe islands difficult: for example, the energy difference between the first unoccupied and occupied levels corresponds to different quantum numbers in the different islands. In addition, if the QD has an odd $N$, we expect to see the same orbital at both negative and positive bias (adding or removing an electron from the same orbital). On the other hand, in QDs with even $N$, electron addition and removal correspond to a different orbital.

\begin{figure*}[t]
    \centering
    \includegraphics[width=\linewidth]{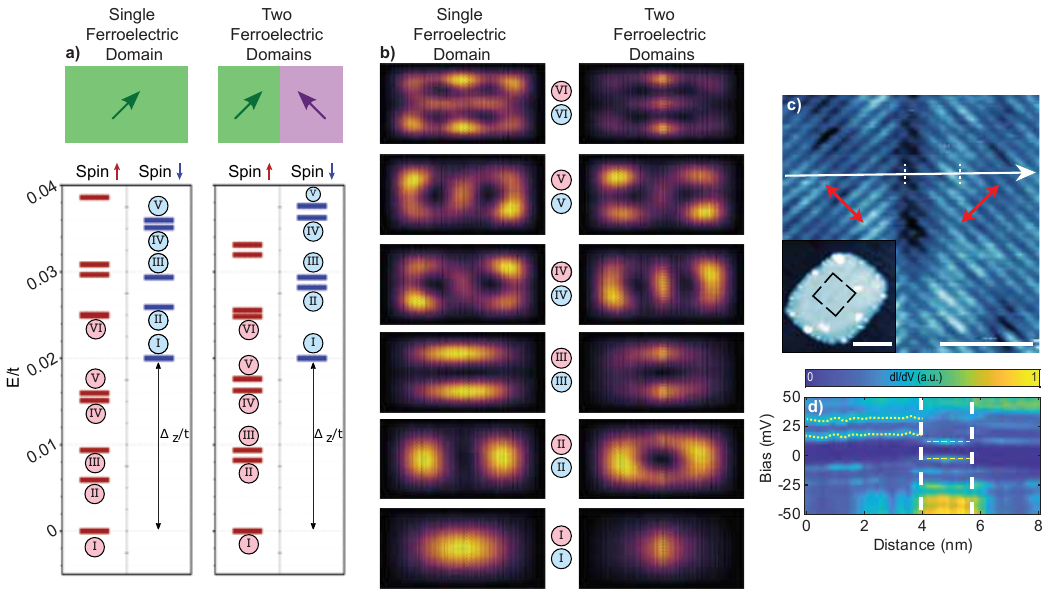}
    \caption{Investigation of multiple domains and low energy gap in SnTe islands. \textbf{a, b.} Calculated energy levels (panel a) and the corresponding LDOS maps (panel b) for two spin-polarized quantum dots of the same size formed in the SnTe island with a single and two ferroelectric domains. \textbf{c.} The boundary between two ferroelectric domains. Red arrows show the alignment of polarization. The white arrow shows the location of the line spectra presented in panel d, with the dashed white lines highlighting specific details presented in panel d (\SI{1}{\volt}, \SI{51}{\pico\ampere}, scale bar \SI{3}{\nano\meter}). The inset shows the SnTe island with two ferroelectric domains. The dashed black line indicates the location of the boundary area presented in panel c (\SI{1.2}{\volt}, \SI{4}{\pico\ampere}, scale bar \SI{10}{\nano\meter}).  \textbf{d.} Line spectra crossing the boundary of two ferroelectric domains. The white dashed lines match the ones in panel c, showing a possible connection between the domain boundary and a narrower energy gap. The resonance separations (yellow dashed lines) are almost equal.}
    \label{fig:fig5}
\end{figure*}

While it is difficult to extract quantitative numbers from QD spectroscopy experiments due to the uncertainty in the level assignment, we consistently observe resonances that have identical spatial patterns. An example of this is shown in Fig. \ref{fig:fig4}b (resonances highlighted by the orange and gray lines). As can be seen, the LDOS patterns show identical spatial modulation, paired with fixed relative intensity. We confirm this pairing by measuring the peak separation in all point spectra taken on the QD (see Fig. \ref{fig:fig4}c for an example). We find that in all cases the separation is 25.6$\pm$\SI{0.6}{\milli\volt}. Considering the particle-in-a-box energy levels, pairs of energy levels having identical spatial pattern must correspond to the same orbital, only with different spin. Then, the energy separation directly gives the strength of the exchange field from the magnetic CrBr$_2$ experienced by the SnTe layer. 
This phenomenon is captured by the theoretical low-energy model for the multiferroic QD presented in Fig.~\ref{fig:fig_theory}c. Results of the calculations with the model in a single-domain island are shown in Figs.~\ref{fig:fig4}d,e (details in the Methods section). It can be observed that the energy levels are spin-polarized, showing a shift due to the Zeeman splitting Fig.~\ref{fig:fig4}d. As a consequence, levels with the same spatial spectral weight appear at different energies, Fig.~\ref{fig:fig4}e, precisely at the one given by the Zeeman splitting $\Delta_z$. 
This theoretical shift is in good agreement with the experimentally observed energy spacing between levels displaying the same spatial distribution.

\subsection{Multidomain multiferroic quantum dot}

Our findings indicate that SnTe islands with a single ferroelectric domain, grown on ML CrBr$_2$, act as multiferroic QDs. This raises the question regarding the SnTe islands with multiple ferroelectric domains: What is the effect of multiple ferroelectric domains on the energy levels of the multiferroic QD? 
To address this, we have performed calculations with the tight-binding model in two islands of the same size with a single domain and two domains (details in the Methods section). 
The results of these calculations are shown in Figs.~\ref{fig:fig5}a,b. 
In both islands, the energy levels are spin-polarized, showing the same shift $\Delta_z$ due to the Zeeman splitting Fig.~\ref{fig:fig5}a. 
A comparison between single and two-domain islands reveals a change in the energy-level spacing, with multiple domains causing increasing confinement and pushing the levels to slightly higher energies. Significantly, the spectral weight in real space associated with these energy levels undergoes a reconstruction due to the ferroelectric domains, see Fig.~\ref{fig:fig5}b. In particular, the spectral weight is concentrated at the domain boundary.

In our experiments, we have analyzed a SnTe island with two ferroelectric domains (Fig. \ref{fig:fig5}c inset).
Figure \ref{fig:fig5}c shows an atomic-resolution scan of the boundary between the two domains.
Two ferroelectric domains can be observed, whose polarization directions are highlighted with red arrows and have been determined by the elongation of the lattice parameter along the ferroelectric distortion.
Line spectroscopy was performed along the white arrow crossing the domain boundary (Fig.~\ref{fig:fig5}d). 
The spectra are different at the ferroelectric domain boundary, with some of the energy levels showing enhanced spectral weight. 
Furthermore, we note that the doubling of the signals (dashed yellow lines in Fig.~\ref{fig:fig5}d) is present here, with a splitting of $\sim$\SI{15}{\milli\volt}. 
Although our experimental results are limited and a more quantitative analysis of the QD levels exceeds the capabilities of our current setup, the reconstruction of the energy levels at the domain boundary in the multiferroic QD is in qualitative agreement with the theoretical prediction of our minimal model.

\section*{Conclusions}

In conclusion, we have demonstrated the emergence of a multiferroic QD in an artificial van der Waals heterostructure combining ferroelectric SnTe and magnetic CrBr$_2$ grown on HOPG. By combining MBE growth with STM and STS measurements, \textit{ab initio} calculations, and low-energy effective models, we show that quantum confinement within the SnTe nanoislands gives rise to discrete electronic states whose spin degeneracy is lifted by the proximity to the underlying magnetic CrBr$_2$ layer. At the same time, the SnTe nanoislands retain their ferroelectric order, bringing together quantum confinement, spin polarization, and ferroelectricity within a single nanoscale system.
Beyond establishing the multiferroic character of the QD, we find that its electronic spectrum is intimately connected to the ferroelectric configuration of the SnTe nanoisland. In particular, the formation of multiple ferroelectric domains modifies the energies and spatial structure of the confined states, demonstrating that the internal ferroic texture provides an additional degree of freedom for engineering the QD spectrum. 

More broadly, our results illustrate how artificial vdW heterostructures can be used to combine distinct functionalities of their constituent layers and transfer them to quantum-confined electronic states. The realization of multiferroic quantum confinement introduces the ferroelectric domain configuration as a potentially reconfigurable control parameter for spin-polarized quantum states. Future developments enabling the controlled manipulation and switching of these domains could therefore provide a route toward non-volatile electrical control of QD states and their spin character. These results establish vdW heterostructure engineering as a promising platform for designing multifunctional QDs and open opportunities for electrically reconfigurable nanoscale spintronic and quantum devices.

\hspace{0pt}

\section*{Methods}

\subsection*{Sample preparation}
All samples were grown in an ultra-high vacuum (UHV) molecular beam epitaxy (MBE)
system incorporated into an STM; thus the sample stayed in UHV for the duration of the experiment. Highly oriented pyrolytic graphite (HOPG) was used as a substrate; it was exfoliated in ambient conditions and subsequently transferred to the preparation chamber with a base pressure of $2.0 \cross 10^{-9}$ mbar, where it was outgassed at \SI{250}{\celsius}. 
CrBr$_2$ was deposited from single-source anhydrous CrBr$_3$ powder using a Knudsen cell, with the temperature of the cell being \SI{350}{\celsius}. CrBr$_3$ was deposited for  40 minutes onto the substrate held at \SI{100}{\celsius}, after which CrBr$_3$ evaporation was shut off and post-growth annealing at \SI{110}{\celsius} took place for 10 minutes. 
After verifying the quality of monolayer CrBr$_2$ via STM, SnTe was deposited for 2 minutes onto the sample at room temperature. SnTe was deposited from a single-source anhydrous powder held at \SI{610}{\celsius} in a Knudsen cell. No post-growth annealing was done. 
Monolayer CrBr$_3$ islands could be grown by increasing the substrate temperature to \SI{160}{\celsius} and above. These growths are discussed in the SI. 

\hspace{0pt}

\subsection*{STM measurements}
All experiments were carried out with a low-temperature (LT) Createc STM system. The base temperature for the measurements was $\sim$\SI{4.2}{\kelvin}. All of the measurements were performed
with Pt/Ir tips dipped into an Au(111) surface. For the spectroscopy measurements presented in Figure 2, we used a conventional lock-in technique at a frequency of 757 Hz with 25 mV modulation. For low-energy spectra presented in Figure 4 and Figure 5, we used a conventional lock-in technique at a frequency of 515 Hz with 1 mV modulation.
\hspace{0pt}

\subsection*{Computational details}

We have performed \textit{ab initio} electronic structure calculations based on density functional theory (DFT) \cite{HK,KS} on a SnTe/CrBr$_2$/Graphene heterostructure.
Calculations were carried out using the plane-wave pseudopotential method as implemented in the \textsc{Quantum ESPRESSO} package \cite{0953-8984-21-39-395502,0953-8984-29-46-465901}. 
The generalized gradient approximation in the Perdew–Burke–Ernzerhof scheme for solids and surfaces (GGA-PBEsol) was employed for the exchange–correlation functional \cite{PhysRevLett.100.136406}. 
On-site Coulomb interactions were introduced through a DFT+U scheme for the treatment of the Cr d orbitals. 
A range of values of U from 5 to 10 eV was tested, leading to the same phenomenology: a high-spin state for Cr and an induced Zeeman spin splitting on the SnTe bands.
Core–valence interactions were described using standard ultrasoft pseudopotentials from the PSLibrary \cite{DALCORSO2014337}.  
The kinetic energy cutoff for the plane-wave basis was set to 70~Ry for the wavefunctions and 600~Ry for the charge density. 
Brillouin zone integrations were performed using a $12 \times 12 \times 1$ Monkhorst–Pack $k$-point mesh. 
A vacuum spacing of 18~\AA\ was included along the out-of-plane direction to avoid spurious interactions between periodic replicas of the vdW heterostructure.

The tight-binding minimal model for the multiferroic quantum dot was solved with pyqula\cite{pyqula}. Explicitly, the Hamiltonian of the model reads
\begin{equation}
\mathcal{H} = \sum_{\langle ij \rangle, \sigma} \left( t + \delta t_{ij} \right) c^{\dagger}_{i\sigma} c^{\phantom{\dagger}}_{j\sigma}
+ \Delta_z \sum_{i,\sigma\sigma'} c^{\dagger}_{i\sigma} s^{z}_{\sigma\sigma'} c^{\phantom{\dagger}}_{i\sigma'} ,
\label{eq:model}
\end{equation}
where $c^{\dagger}_{i\sigma}$ creates an electron with spin $\sigma$ in the Sn orbital at site $i$ of the square lattice, $\langle ij \rangle$ denotes nearest-neighbor pairs, and $s^{z}$ is the spin Pauli matrix. The first term describes the electron-doped SnTe conduction band, whose ferroelectric distortion enters through the bond modulation $\delta t_{ij}=\pm\delta t$, with the pattern of signs set by the polarization direction of the domain containing the bond (Fig.~\ref{fig:fig_theory}c). The second term is the exchange-induced Zeeman splitting produced by the underlying CrBr$_2$ layer. The monodomain island calculations were performed in a $17\times27$ sites rectangular dot, while the two-domain island was simulated in a $34\times27$ sites dot with a boundary at the center of the larger side. 
The sign of $\delta t$ was flipped in the left and right areas to simulate the ferroelectric domains.
Values of $\delta t=\pm0.3t$ and $\Delta_z=0.01t$ were considered in all the simulations. 

\hspace{0pt}

\section*{Data availability}
The data that support the findings of this study are available from the corresponding author upon reasonable request.

\hspace{0pt}

\section*{Acknowledgements}
We thank Dr.~Aleš Cahlík for fruitful discussion. This research made use of the Aalto Nanomicroscopy Center (Aalto NMC) facilities and was supported by the Research Council of Finland (Academy Research Fellow funding nos.~338478, 359922, 347266, 368478, 369367, project nos.~370910, 370911, 370912, the Finnish Centre of Excellence in Quantum Materials QMAT no.~374166, and the Finnish Quantum Flagship project no.~358877), the European Research Council (ERC StG TITAN no.~101039500, ERC AdG GETREAL no.~101142364, and ERC CoG ULTRATWISTROICS no.~101170477). We acknowledge the financial support of the Finnish Ministry of Education and Culture through the Quantum Doctoral Education Pilot Program (QDOC VN/3137/2024-OKM-4) and the computational resources provided by the Aalto Science-IT project. 

\hspace{0pt}

\section*{Author contributions}

Antti Karjasilta: Validation, Investigation, Formal analysis, Writing - Original Draft. Mohammad Amini: Conceptualization, Investigation, Methodology. Liwei Jing: Investigation. Robert Drost: Funding acquisition, Validation, Visualization. Shawulienu Kezilebieke: Conceptualization, Supervision, Funding acquisition. Jose Lado: Conceptualization, Supervision, Funding acquisition. Peter Liljeroth: Conceptualization, Supervision, Funding acquisition, Writing - Review and Editing. Adolfo O. Fumega: Conceptualization, Funding acquisition, Methodology, Validation, Investigation, Formal analysis, Visualization, Writing - Original  Draft. 

\hspace{0pt}

\section*{Competing interests}
The authors declare no competing interests.

\phantomsection
\addcontentsline{toc}{section}{\refname}
\bibliography{references.bib}

\end{document}